\documentstyle{elsart}
\input{psfig}

\begin{document}
\begin{frontmatter}
\title{Measurement of the atmospheric neutrino flavour composition in Soudan 2}
\author[3]{W.W.M. Allison}, 
\author[4]{G.J. Alner},
\author[1]{D.S. Ayres}, 
\author[6]{W.L. Barrett}, 
\author[2]{C. Bode},
\author[2]{P.M. Border}, 
\author[3]{C.B. Brooks}, 
\author[3]{J.H. Cobb}, 
\author[4]{D.J.A. Cockerill}, 
\author[4]{R.J. Cotton},
\author[2]{H. Courant}, 
\author[2]{D.M. DeMuth},
\author[1]{T.H. Fields},
\author[1,2,3]{H.R. Gallagher},
\author[4]{C. Garcia-Garcia\thanksref{b}}, 
\author[1]{M.C. Goodman}, 
\author[2]{R.N. Gray,} 
\author[2]{K. Johns\thanksref{c}},
\author[5]{T. Kafka},
\author[2]{S.M.S. Kasahara}, 
\author[5]{W. Leeson},
\author[4]{P.J. Litchfield},
\author[2]{N.P. Longley\thanksref{d}},
\author[2]{M.J. Lowe\thanksref{e}},
\author[5]{W.A. Mann}, 
\author[2]{M.L. Marshak}, 
\author[1]{E.N. May},
\author[5]{R.H. Milburn}, 
\author[2]{W.H. Miller}, 
\author[2]{L. Mualem},
\author[5]{A. Napier}, 
\author[5]{W. Oliver}, 
\author[4]{G.F. Pearce}, 
\author[3]{D.H. Perkins}, 
\author[2]{E.A. Peterson}, 
\author[3]{D.A. Petyt},
\author[1]{L.E. Price}, 
\author[2]{D.M. Roback\thanksref{f}},
\author[2]{K. Ruddick},
\author[2]{D.J. Schmid\thanksref{h}}, 
\author[5]{J. Schneps}, 
\author[2]{M.H. Schub},
\author[1]{R.V. Seidlein},
\author[2]{M.A. Shupe\thanksref{c}},
\author[3]{A. Stassinakis},
\author[5]{N. Sundaralingam\thanksref{i}},
\author[3]{J.Thomas},
\author[1]{J.L. Thron},
\author[2]{V. Vassiliev}, 
\author[2]{G. Villaume}, 
\author[2]{S.P. Wakely}, 
\author[5]{D. Wall}, 
\author[2]{S.J. Werkema\thanksref{k}},
\author[3]{N. West},       
\author[3]{U.M. Wielgosz}                 
\address[1]{Argonne National Laboratory, Argonne, IL
60439, USA }
\address[2]  {University of Minnesota, Minneapolis, MN
55455, USA }
\address[3]{ Department of Physics, University of Oxford, 
Oxford OX1 3RH, UK }
\address[4]  {Rutherford Appleton Laboratory, Chilton, Didcot,
Oxfordshire OX11 0QX, UK }
\address[5]{ Tufts University, Medford, MA 02155, USA }
\address[6]  { Western Washington University, Bellingham, WA 98225, USA }

\thanks[b]  {Now at IFIC, E-46100 Burjassot, Valencia, Spain  }
\thanks[c]  {Now at University of Arizona, Physics Department, Tucson, AZ 85721, USA}
\thanks[d] {Now at Swarthmore College, Swarthmore, PA 19081 USA }
\thanks[e] {Now at Dept of Medical Physics,University of Wisconsin, Madison,
WI 53705, USA}
\thanks[f] {Now at the Dept of Radiology, University of Minnesota, 
Minneapolis, MN 55455, USA}
\thanks[h] {Now at Kodak Health Imaging Systems, Dallas, TX, USA}
\thanks[i] {Now at Edward Waters College, Jacksonville, FL 32209, USA}
\thanks[k] {Now at Fermi National Accelerator Laboratory, 
Batavia, IL 60510, USA }

\begin{abstract}
The atmospheric neutrino flavour ratio measured using a 1.52 kton-year 
exposure of Soudan~2 is found to be 
$0.72\pm0.19^{+0.05}_{-0.07}$ relative to the expected value from a Monte
Carlo calculation.  
The possible background of interactions of
neutrons and photons produced in muon interactions in the rock surrounding
the detector has been investigated and is shown not to produce low values of 
the ratio. 
\end{abstract}
\end{frontmatter}

\section{Introduction}
The flavour content of atmospheric neutrino interactions has previously
been measured
in four underground experiments \cite{kam,imb,frejus,nusex}.
The first two experiments, those performed in water Cherenkov detectors, found
that the ratio of $\nu_\mu$ to $\nu_e$ events (tagged by the outgoing lepton)
was different from that expected from their Monte Carlo calculations. 
On the other hand
the latter two experiments, carried out in iron calorimeters, found consistency
, albeit with inferior statistical
precision.  
\par In order to cancel the uncertainties in the overall cosmic
ray flux it is desirable to present the result in the form of the 
double ratio
\begin{displaymath}
R_t=
\frac{\left(\frac{\nu_\mu}{\nu_e}\right)_{data}}
{\left(\frac{\nu_\mu}{\nu_e}\right)_{MC}} \,.
\end{displaymath}
The water
Cherenkov experiments have selected as a measure of the $\nu_\mu$ rate,
single track (muon) events and the $\nu_e$ rate, single shower
(electron) events. One can then form the experimental ratio
\begin{displaymath}
R=
\frac{\left(\frac{tracks}{showers}\right)_{data}}
{\left(\frac{tracks}{showers}\right)_{MC}} \,.
\end{displaymath}
  The water Cherenkov detectors found values of $R$ between 
$0.54 \pm 0.07$ and $0.62 \pm 0.08$ \cite{Kamneutron}. 
The Frejus iron calorimeter 
experiment, using all events rather than only single prong events and
including uncontained events, found a double ratio consistent with 1.0
\par Since the first reports of this anomaly much effort has been invested in
verification of the Monte Carlo calculations \cite{Gaisser} and in checking
the experimental procedures \cite{Kamtest}.
No convincing explanation for the water Cherenkov anomaly
not involving new physics has been put forward.  However 
there still may be undetected backgrounds or experimental problems.  
In particular
it has been postulated \cite{Ryaz} that the effect may be due to a
background of neutron produced events, though evidence against this has been 
produced by the Kamiokande experiment \cite{Kamneutron}.  
The possibility remains that the flavour content has changed 
between the $\nu$ production in the upper atmosphere and their
interaction in the underground detectors, implying that $\nu$ flavour 
oscillations have taken place, that neutrinos have mass and that physics beyond
the standard model is being observed.
\par In this letter we report a measurement of the flavour ratio in Soudan~2
from an exposure of 1.52 fiducial kton-years.  The value of $R$ obtained is:
\par  $R=0.72\pm0.19 ^{+0.05}_{-0.07}$ 
\par Although on its own the deviation from unity 
is not  significant,  
the agreement of the sign of the discrepancy  with
the water Cherenkov data adds weight to the hypothesis of a real effect.
\par Soudan 2
is an iron calorimeter with 
different experimental systematics from the water Cherenkov detectors
and with a different geometry and detection
technique to the Frejus experiment.  
Background events produced by neutral particles
entering the detector from the interactions of cosmic ray muons in the 
surrounding rock are tagged by a hermetic active shield. 
We show that our low value of $R$ is not due to a 
contamination from such events.  Our measured value of the track/shower ratio 
for neutron produced events does not support
the hypothesis that the anomaly in the Kamiokande and IMB
experiments is due to such a contamination.

\section{The Soudan 2 detector}

The Soudan 2 experiment is located 710 meters underground in the 
Soudan Underground
Mine State Park, Soudan, Minnesota,USA.  The main detector is a  
time projection,tracking 
calorimeter with a total mass of 963 metric tons. It consists of 224  
modules each weighing 4.3  tons and having an average density of 1.6
g/cc.  It is surrounded by an 
active shield of aluminum proportional tubes.  
\par About 85\% of the mass of a module 
is provided by 1.6 mm thick sheets of corrugated 
steel.  The sheets are stacked to form a hexagonal `honeycomb' structure.  
Plastic
drift tubes (1.0 m long and 15 mm in diameter) fill the spaces
in the honeycomb.
An 85\% argon/15\% CO$_{2}$ gas mixture is recirculated through the modules.
Ionization deposited in the gas 
drifts toward the closer end of 
the tube in an 180 volt/cm electric field.
The drift velocity is approximately $0.6$ cm/$\mu$sec, 
which yields a 
maximum drift time of 83 $\mu$sec.  
\par On reaching the end of the tube, the 
charge is detected by vertical anode wires and horizontal 
cathode strips.  The signals from widely separated wires and strips are  
summed to reduce the number
of readout channels. The summing is designed
such that matching a pulse from 
an anode and cathode channel uniquely identifies
the module and tube from which the ionization drifted.
The 
signals are digitized every 200 nsec and are stored in a 1024 word buffer.
The 
primary trigger condition requires activity at different times in any 7 anode
OR 8 cathode channels out of any block of 16 channels within a total gate
of  72 $\mu$sec.
Further details of module construction may be found in reference 
\cite{modcon}, its performance in a cosmic ray
test stand in \cite{modperf} and the performance at the Soudan mine in 
\cite{gallagher}.
               
\par The calorimeter is surrounded by a 1700 m$^2$ active shield 
designed to identify particles which enter or exit the 
detector cavern.  The  shield covers about 97\% of the total solid
angle.
 The basic element is an extruded aluminium manifold, up to 8m long,
consisting of eight hexagonal proportional tubes
arranged in two layers of four.  
The four tubes in each layer are connected together and 
read out as one signal.  The random rate in a tube layer coming from
natural radioactivity ($\sim300$~Hz~m$^{-2}$) would produce an unacceptably high
rejection rate. Thus a coincidence of an adjacent 
inner and outer layer is required to signal a high energy
particle entering or leaving the cavern.  The measured efficiency of a 
coincidence for a single, high energy particle traversing a shield element
is 95\%.
More details of the shield construction
and performance can be found in reference \cite{shield}.

 The completed detector runs at a trigger rate of $\approx$ 0.5 
Hz. Approximately two thirds of triggers come from cosmic ray muons passing 
through the detector.  Most of the remainder are due to electrical noise
or naturally occurring radioactivity.
The detector routinely runs with an overall efficiency of $\sim80$\% which
rises to over $90\%$ during nights and weekends when the laboratory is not
occupied.
Immediately after completion of a run the data are processed to
reconstruct the events and sort them into output files of candidate 
events for various physics analyses.

 Every 240 seconds a data acquisition sequence is initiated, 
irrespective of detector activity.  These `pulser' events provide a snapshot 
of the background levels in the main detector and 
are used as underlying events to add detector noise to 
Monte Carlo events.  

\section{Data Analysis}
\subsection{Data reduction}
\label{sec:dataanal}
    The data considered in this letter come from a 1.52 kton-year exposure
between April 1989 and December 1993.  During this
period the detector was under construction, starting with a total mass of
275 tons and ending with the complete 963 tons.  A total of 43 million
triggers was taken.

  The goal of the data reduction is to obtain a sample of
`contained events' which will be used both for the atmospheric neutrino
analysis described here and for a search for proton decay.  A contained
event is defined as one in which no primary particle in the event 
leaves the fiducial volume of the detector, defined by a 20 cm depth cut on
all sides of the detector.

\par The events are passed through a
software filter to reject events with tracks entering or leaving the
fiducial volume
(mostly cosmic ray muons) or events which have the characteristics
of radioactive background or electronic noise.  Approximately 1 event
per 1500 triggers passes this filter.
\par  The selected  events are then double scanned to check containment and 
to reject background events, using an  interactive graphics program.  
The
main backgrounds  are residual radioactive and electronic noise,
badly reconstructed cosmic ray muons
and  events where muons
pass down the gaps between individual modules, either finally entering a
module and stopping or interacting in material 
in the gap and sending secondary tracks
into the modules.  Any event with a track which starts or ends on a
gap, or which can be projected through
a gap to the exterior of the detector is rejected. In addition, events with
a vertex in the crack region are rejected.
Differences between scanners are resolved by a second level scan.
Approximately 1 event in 40 passed by the program filter is finally
selected as contained.
The average efficiency of individual scanners in selecting contained events
was 93.5\%.  Further details of the event selection procedure can be found
in reference \cite{gallagher}.

\subsection{Monte Carlo analysis}
\par A Monte Carlo simulation of the experiment has been developed
which reproduces as closely as possible the experimental data.  In particular,
 Monte Carlo events have been made visually
indistinguishable from true data events
to experienced physicist scanners.  This currently
enables Monte Carlo events to be inserted randomly 
into the data stream 
and to be processed simultaneously with the data events, 
ensuring that they are treated identically.
This version of the Monte Carlo program was not available at
the beginning of the experiment.  The last third of the 
data set reported here had
Monte Carlo events inserted at the scanning level. The first 
two thirds were initially processed independently of the Monte Carlo.
Although 
the Monte Carlo corresponding to this earlier data was processed and scanned 
separately great care was taken to follow the same procedures as for the
real data and thus avoid biases.
\par Monte Carlo events equivalent to 5.9 times the exposure of the real data 
were generated and passed
through exactly the same data analysis procedure as described in section
\ref{sec:dataanal}.
\par The neutrinos were generated using the BGS flux\cite{barr}. 
The variation of the $\nu$ intensity with the solar cycle was corrected
using neutron monitor data\cite{gallagher,beiber}.
\par At the low $\nu$ energies characteristic of the atmospheric flux the
predominant interactions are quasi-elastic or resonance production.  Full
details of the event generation process and a detailed comparison with
all available low energy data are given in reference
\cite{gallagher}.  Nuclear physics effects were represented by the Fermi gas
model.  Rescattering of pions within the nucleus was applied
using data obtained by comparison of bubble chamber $\nu$
interactions on deuterium and neon \cite{intranuke}.
\par Events were generated to simulate the exact
size and configuration of the detector as it grew during this exposure.
Particles produced in
the neutrino interactions were tracked through the detector geometry using
the EGS  and GEISHA  codes.  
Particles crossing the drift tubes had amounts
of ionization deposited in the gas selected from the distribution of
reference \cite{allison}.  The
ionization was drifted, with appropriate attenuation and diffusion, to the
anode wires where the effects of the 
avalanche and electronics response were closely simulated. 
The generated
event was superimposed on a pulser trigger which
reproduces noise and background in the detector as 
they vary with calendar time.
\par A comparison  of physical quantities, including
topologies and energy distributions, between data and Monte Carlo
showed no discrepancies outside the
possible effects of the atmospheric neutrino flavour anomaly discussed in 
this paper.  In addition the Monte Carlo representation of tracks and showers
has been tested against data taken at the Rutherford Appleton Laboratory ISIS
test beam facility which provided electrons, pions and muons
 up to a momentum of 400 MeV/c and protons up to 800 Mev/c.
\subsection{Event classification and reconstruction}
\par  The aim of this analysis is to measure the flavour content of
neutrinos incident on the detector after their
passage from the upper atmosphere. Given the 
predominance of 
quasi-elastic scattering the relative rate of single shower (electron)
and single track (muon) events is a good measurement of the flavour
content.  It is also the measurement made in the water Cherenkov detectors.
We expect in the future to use the superior track separation
and reconstruction properties of Soudan 2 to flavour classify events with
multiple tracks but the objective of this paper is to repeat
the earlier measurements.
\par  The lepton flavour of each event is determined by the second level
scanners who flag them as `track', `shower' or `multiprong'.  Tracks
which have heavy ionization and are straight are further classified as
`protons'.  Proton recoils accompanying tracks and showers are an additional
tag of quasi-elastic scattering and are ignored in the classification.  Any
second track or shower in the event results in a multiprong 
classification.  As a test of the systematic uncertainties introduced by
the classification process, all scanning was done independently by two groups 
prior to merging for the final results.
\par  The quality
of the flavour assignment was measured using the Monte Carlo data.
Table \ref{mismatrix} gives the
identification matrix for Monte Carlo events selected as contained.
\begin{table}[h]
\caption[Monte Carlo identification matrix]
{Monte Carlo identification matrix.\\}
\label{mismatrix}
\begin{tabular}{|l|cccc|}
\hline
       & &\multicolumn{2}{c}{Assigned}&\\
Generated    & Track & Shower & Multiprong & Proton \\
\cline{1-5}
$\nu_\mu$ cc       & 242 &  3 & 98 &  6  \\
$\nu_e$ cc         & 15 &255 & 110 &  1  \\
Neutral current    & 21 &  9 & 44 & 18  \\
\hline
\end{tabular}
\end{table}

It can be seen that 87\% of events assigned as
tracks have muon flavour and 96\% of showers
electron flavour.  The identification matrix is consistent between the
first two thirds of the data when the scanners were aware that they
were scanning MC events and the last third when the events were
randomly mixed.  The ratio of accepted muon to electron charged current 
events is 
approximately 1:1, different from the expected ratio of 2:1 from the 
$\pi \rightarrow \mu \rightarrow e$ decay chain. At these low
energies threshold effects due to the difference in the muon and electron
masses cause the generated event 
ratio to be approximately
1.5:1.  Acceptance differences for high energy muons and electrons and
the cuts required to remove background produced by cosmic ray muons 
passing down the gaps between modules further
reduce the ratio.
\par   
Each contained event is reconstructed, using the interactive graphics system,
to determine its position and the energy of the identified particles.  
A vertex is assigned, and the
location of the ends of any tracks is marked.

\par Electron showers are reconstructed using a clustering algorithm to 
select all hits lying within 60 cm of their nearest 
neighbour.  The shower
direction is determined using these hits and the vertex defined by the scanner.
The shower energy is calculated from the number
of hits.  The energy is calibrated using the results of a test beam exposure 
of a module  to electrons below 400 MeV
at the Rutherford
Appleton Laboratory ISIS facility  and
to Monte Carlo showers at higher energies. 
Early data had some contamination of the shower sample from electrical
breakdown in some modules.  This was much improved as the experiment
progressed by optimization 
of the wireplane  voltages and refurbishment of the worst
modules.  In order to remove this contamination a cut which required 
$\ge9$ hits was applied to the showers, corresponding to an energy cut of
approximately 150 MeV.  Raising this cut 
had no significant effect on the ratio $R$.    

\par Tracks are reconstructed by fitting a polynomial 
to the hits belonging to the track. The amount of material traversed by 
the particle along the fit trajectory is calculated by tracking the polynomial
through the detailed geometry of the module. The range 
is then converted into 
a particle energy by integrating the Bethe-Bloch equation, assuming a muon 
mass.  
The energy calibration has again been checked using data from the test beam
exposure.
A minimum of
6 hits on the track was required, corresponding to a muon 
kinetic energy cut-off of approximately 40 MeV.
Tracks produce a very regular pattern of hits in the honeycomb geometry which
breakdown processes do not reproduce and
there is no evidence of such
contamination
of the track sample.

\subsection{Shield data and the identification of $\nu$ events}
\label{sec:shield}
   A total of 723 data events are classified as contained. 
This is much greater than the expected neutrino rate of about 100 
events/kton-year.
We conclude that the majority of these events 
are due to the interactions of neutral particles (neutrons or photons)
produced by muon interactions in the rock around the detector.
The active shield
is designed to flag such events by detecting the muon and/or other charged
particles which are produced in the muon interaction but do not enter the
main detector.  
It was placed as close to
the cavern wall and as far away from the detector as possible to maximize
the probability
of detecting the accompanying charged particles.  Calculations
\cite{pdk642} indicate that only a few per cent 
of such events will not have charged
particles traversing the shield.  

\begin{figure}                                             
\leavevmode\psfig{file=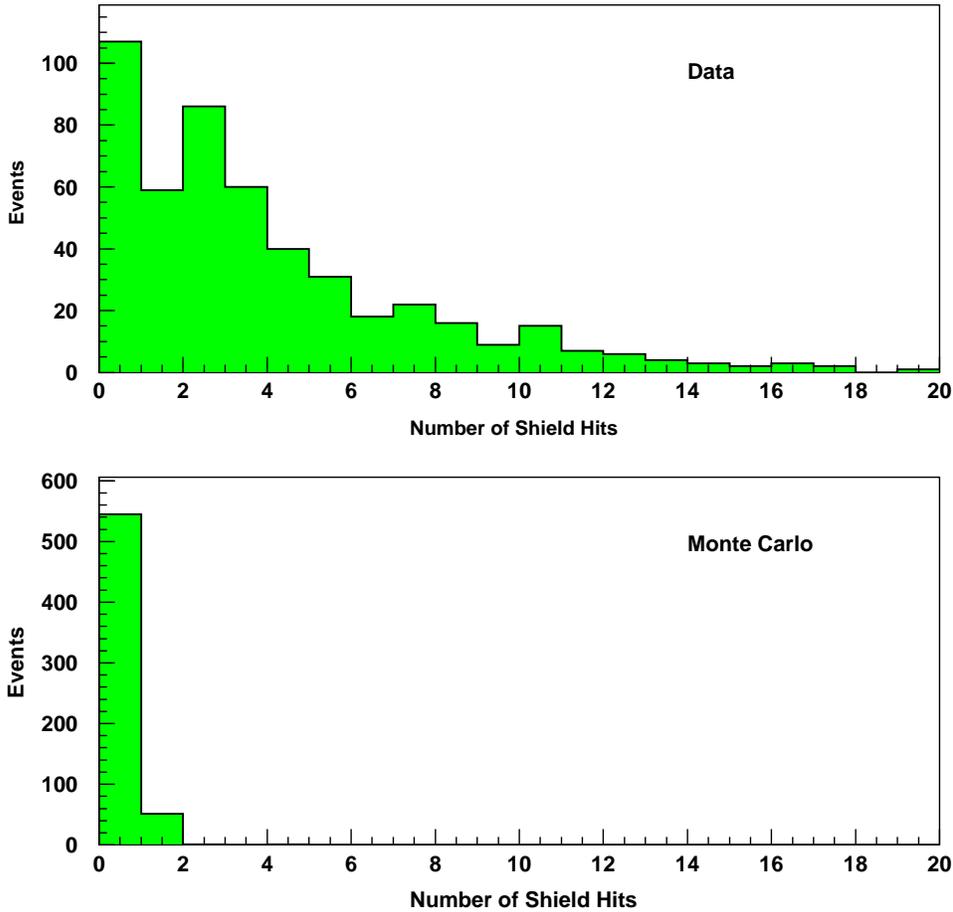,width=5.5in}
\caption
{\label{adjc}
Histogram of the number of shield hits for the data events (top) and Monte 
Carlo events (bottom).  The data is a mixture of neutrino events and rock
background.  The Monte Carlo plot contains only generated neutrino events.}
\end{figure}

\par  Figure \ref{adjc}(top) is a histogram of the number of coincident
shield hits accompanying each track or shower event.  The events with no shield
hits are defined as our $\nu$ sample (`gold' events) and the events with
shield hits are defined as arising from muon interactions (`rock' events).
\par   Figure \ref{adjc}(bottom) shows the same plot for Monte Carlo 
contained events.
The events
with shield hits are due to random shield hits in the background
pulser events during the allowed time window of the MC event. 
A total of 53 out of 598 (8.9\%) of Monte Carlo events had random
shield coincidences.  
The random  veto events are almost all of multiplicity 1, consistent with
the veto being due to Compton electrons produced by photons
from the natural radioactivity in the rock.  The
random vetoing of real events is simulated by selecting only
events with no shield hits for the Monte Carlo gold sample.

\par  Rock events produced by a muon which passes through the cavern should
give at least two shield hits.
The one shield hit events are a combination of zero shield hit events with
random hits, genuine one shield hit events where the entering charged
particle is stopped in the cavern and potential two shield hit events 
with a missing  hit due to shield inefficiency.  The efficiency of the shield
has been measured using cosmic ray muons detected in the main detector. It 
ranges from 81\% during the early data runs before the geometrical coverage 
was complete
to 93\% at the end of this data period, equal to the convolution of the
geometrical coverage and the single tube efficiency.
 Using the number of 0, 1 and 2 hit
events we estimate that $7\pm2$ gold events are due to muon interactions
with a charged particle passing through the shield which was not recorded
due to shield inefficiency.
\par Our sample of rock events,  used to determine
the properties of any potential non-neutrino background, was defined as
those with $\ge2$
shield hits since the one shield hit event sample also contains randomly vetoed
neutrino events 
\par  Table \ref{raw_numbers} gives the raw numbers of gold, rock and gold 
MC events
in our sample, divided  into track, shower, multiprong and proton.  The ratio
of single prong to multiprong events in our data is higher than in previous
experiments.  Our track energy threshold is lower than the Cherenkov
threshold in water and both track and shower thresholds are considerably
lower than those of Frejus.  Also the requirement that no track ends on a
gap between modules preferentially rejects multiprongs.

\begin{table}[h]
\caption[Classifications for the contained events
before corrections. ]{Classifications for the contained events 
before corrections.\\}
\label{raw_numbers}
\begin{tabular}{|l|c|c|c|c|}
\hline
                     & Track & Shower & Multiprong & Proton \\ \hline 
Data: gold           &   47  &   60   &  51 & 10 \\
Data: rock           &  160  &  169   &  90 & 56 \\
MC                   &  278  &  267   & 252 & 25 \\
\hline
\end{tabular}
\end{table}

\section{Measurement of the flavour ratio}

\subsection{Background determination}
                                         
\par In section \ref{sec:shield} it was estimated that a background
of $7\pm2$ rock events was expected in the gold sample 
because of shield inefficiency.  
There is also the possibility
that neutrons or photons may enter the detector without being accompanied
by charged particles in the shield.  
Our large sample of rock events enables
us to investigate this potential background
by studying the depth
distribution of the events in the detector. 
\par
\begin{figure}
\leavevmode\psfig{file=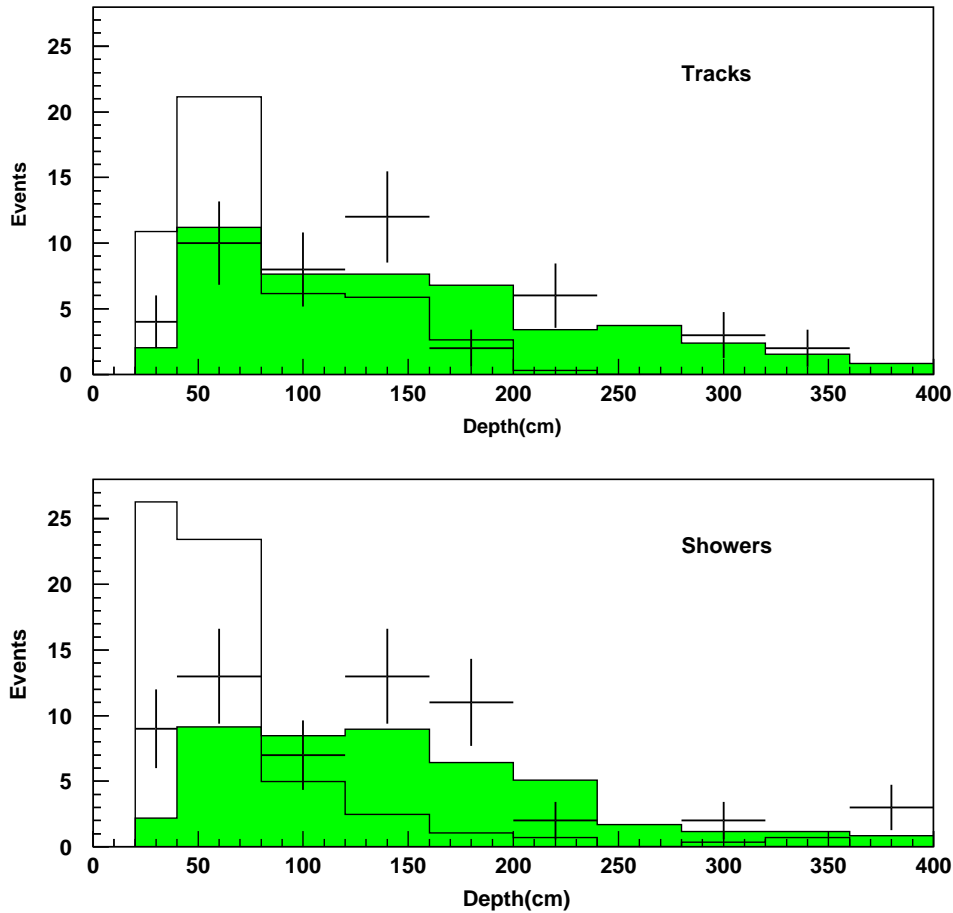,width=5.5in}
\caption
{\label{depth}
The depth distributions for tracks (top)  and showers
(bottom).
The data points are the gold data, the shaded histogram is
the gold Monte Carlo, normalized to the experiment exposure, 
and the unshaded histogram
is the rock data, normalized to the same number of events as the data sample.}
\end{figure}

  The
events produced by photons and neutrons will be attenuated towards the centre
of the detector,
whilst the neutrino
events will be uniformly distributed through the detector.
 Since the directions of the particles produced in neutron interactions
will not in general be
the same as that of the incident neutron we cannot directly measure the
distance that the neutron travelled through the detector.  Instead
we define a measure of the 
proximity of the event to the detector exterior by calculating
the minimum perpendicular distance from the event vertex to the detector edge. 
Since few rock photons and neutrons are expected to travel upwards and
the base of the detector  does not have an excess
of rock vertices, the floor is not considered to be an `edge' for the 
purposes of this calculation.  Figure \ref{depth} shows this 
depth distribution for gold, 
Monte Carlo and rock tracks and showers. 
 The Monte Carlo distributions
are normalized to the exposure of the experiment and the rock sample
is normalized to the same number of events as the data sample.

\par
\begin{figure}
\leavevmode\psfig{file=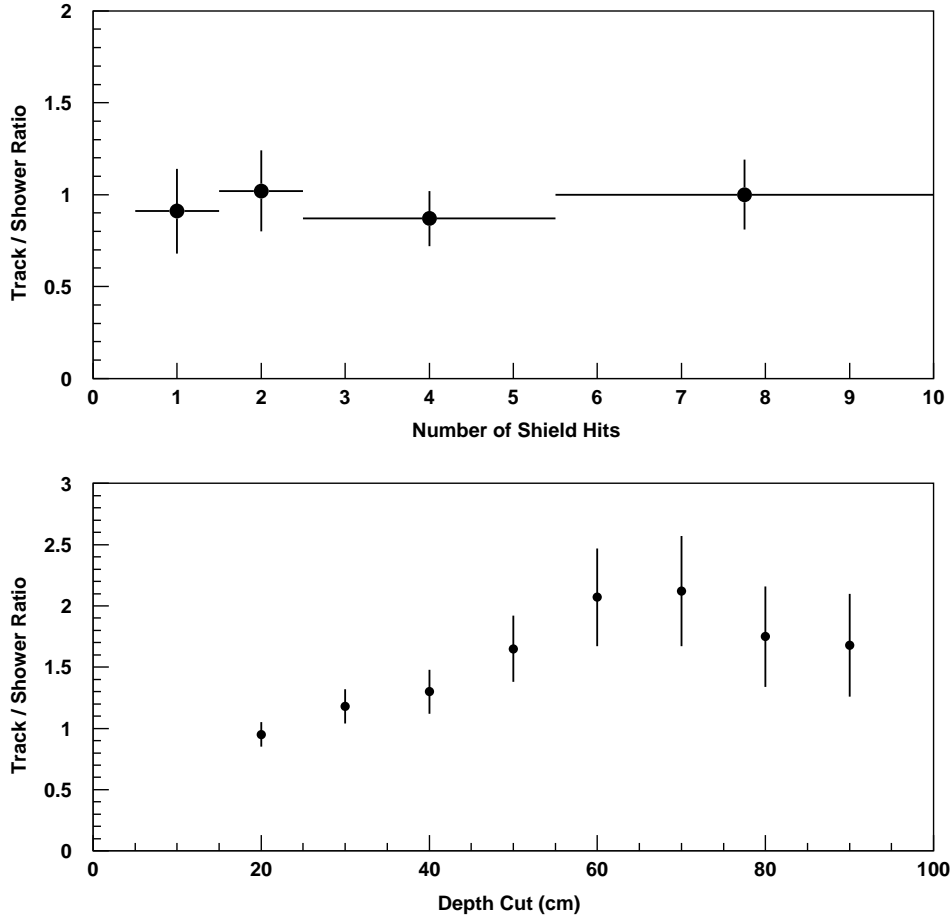,width=5.5in}
\caption
{\label{ratio}
The track/shower ratio for rock events 
as a function of the number of shield hits (top)
and the depth cut (bottom).}
\end{figure}

  The rock shower and rock track depth 
distributions are
different.  The track distribution is consistent with being produced by
incoming neutrons with an interaction length of approximately 
80 cm.  The shower distribution appears to have two components, 
a long range component consistent with neutrons and a short range component
which we attribute to photons.  The short range component has a depth
distribution consistent with the photon conversion length of 15cm measured in
a module at the ISIS test beam \cite{garcia}.  
Figure \ref{ratio}(bottom) shows the integral 
track/shower ratio as a function of 
depth cut.  The ratio rises as events near the edge of the detector are
removed, reaching a plateau at a depth cut of around 60 cm when the photon
component has been fully attenuated.

\par Comparison of the Monte Carlo  and the gold data
depth distributions indicates that there may be a small excess of events
at small depth in the shower sample whilst the track distribution closely
follows the expected neutrino distribution.  However the discrimination
between rock and MC distributions is better in the shower sample because
of the short distance photon component.  

\par  The interactions of the neutral particles in the detector are not 
expected  to be strongly correlated with the number of shield hits.  This is 
verified in figure \ref{ratio}(top) which shows the track/shower ratio to be
constant as a function of number of shield hits.  We therefore use the full
rock sample to estimate the amount of zero shield hit rock background in the
gold sample.
We have made $\chi^2$
fits of the shape of the gold distributions of figure \ref{depth}
to the sum of the shapes of the
rock and MC distributions with one free parameter, the fraction of the rock
background required in the gold distribution.  The bins at large depths were
combined to give at least 5 gold events per bin, giving a total of 6 bins
in each distribution.
A fit with the background set to zero gave $\frac{\chi^2}{NDF}$ of 0.22 and 1.42
for the track and shower distributions respectively.
The fit of the tracks is very good
while the shower fit probability is about 20\%.  To this level the fits do not
require any background contribution.  However as described above we expect 
some  rock contamination so we continue with a second fit
which allows a free amount of background
in each distribution.  The $\frac{\chi^2}{NDF}$ are now 0.19 and 0.54.
The track fit is not improved but there is
a significant drop in $\frac{\chi^2}{NDF}$ for the shower distribution.  
As expected
from examination of figure \ref{depth}  more background is suggested in
the shower depth distribution than in the track distribution.  
We find $4.5\pm6.9$ background events  in the
track sample and $14.2\pm5.9$ in the shower sample, yielding a
track/shower ratio of $0.3 \pm 1.4$ for background events.
This is consistent
with, but numerically rather different to the measured rock event ratio
of $0.95 \pm 0.10$.  
In a third fit we constrain the ratio of the 
background in the gold tracks and showers to be equal to the measured ratio
in the rock events.  The $\frac{\chi^2}{NDF}$ for the combined track and shower fit
is 0.42 and it gives a total background of $20.6\pm8.9$ events, which are
to be divided between tracks and showers in the
measured rock ratio.
 The number of background
events is consistent with those found in the unconstrained fits
 and the fit quality is equally good. We use the constrained fit  
in the calculation of $R$ since
 it is consistent with the other fits,
 it uses the maximum amount of measured information,
and it produces the smallest errors on $R$.
The small systematic errors which are
introduced by the assumption that the background in the gold sample is
represented by the rock sample are considered in the next section.
       
\subsection{Calculation of $R$}
\par  To calculate $R$ we correct the raw numbers of gold events using the
background estimated in the constrained fit.  Note that the error on the
correction depends on the errors on the fraction of rock events in the
gold sample and the
measured rock ratio, not on the uncorrelated errors on the number of
background events.
The numbers entering the calculation and the corrected and uncorrected
values of $R$ are given in table \ref{results}.  The error on $R$ includes the
error due to the background subtraction as well as the statistical errors
on the numbers of data and Monte Carlo events. The
background correction has only a small effect on the value of $R$ but adds to
the error.

\begin{table}
\caption[]{ Values of the various quantities used in the calculation
of $R$.  The Monte Carlo numbers in parentheses are scaled by the nominal factor
of 5.9.\\}
\label{results}
\begin{tabular}{|l|l|}
\hline
  Number of gold tracks      &   47  \\
  Number of gold showers     &   60  \\
  Number of MC tracks        &  278 (47.1)\\
  Number of MC showers       &  267 (45.3)\\
  Number of rock tracks      &  160\\
  Number of rock showers     &  169\\
  Rock track/shower ratio    &  $0.95\pm0.10$ \\
  Fraction of rock events in gold sample & $0.062\pm0.027$\\
  Corrected number of $\nu$ tracks & 37.0\\
  Corrected number of $\nu$ showers & 49.4\\
 & \\
\hline
  Raw value of $R$ (no background correction) &$ 0.75\pm0.16$\\
\hline
  Corrected value of $R$ & $0.72\pm0.19$\\
\hline
\end{tabular}
\end{table}

\par The systematic errors which could effect the value
of $R$ may be divided into the following categories:
\begin{itemize}
\item Systematic uncertainties in the incident neutrino flux ratio. A number
of calculations have been made of the neutrino fluxes, summarized and
corrected in a recent review \cite{Gaisser}.  There is agreement that
although the absolute rate is uncertain to the order of $\pm20\%$ the flux
ratio is  much better known.
We take an
uncertainty of $\pm5\%$ in $\frac{\delta R}{R}$.
\item Systematic uncertainties in the neutrino generator.  These include 
factors such as the uncertainty in the axial vector mass, the various cross
sections, the treatment of Fermi motion, the uncertainty in the intranuclear
absorption etc.  All these factors are considered in more detail in
reference \cite{gallagher}.  It should be remembered that neutrino 
universality constrains the $\nu_e$ and $\nu_\mu$ cross sections to be equal
up to mass effects.
We estimate they contribute an amount $\pm0.03$
in $\delta R$.
\item Systematic uncertainties introduced by the scanning process.  In order 
to estimate this contribution the data was independently scanned and 
classified by two different groups before the groups merged.
A value of $R$ was calculated by each group independently.
The difference in the raw $R$ value was 0.02.
Since
the final result was obtained by combining the two groups and resolving
differences we expect the final
error due to systematic scanning differences to be smaller than this.  
However we take the full difference and assign a systematic error on $R$ of
$\pm0.02$.
\item Systematic uncertainties on the background subtraction.  
The main systematic 
error lies in
the assumption that the 
track/shower ratio of the zero shield hit rock background is the same as
that of the $\ge2$ shield hit rock events. It was shown in 
figure \ref{ratio}(top) that this ratio is constant as a function of number
of shield hits.
However, it might be expected that zero shield hit events 
arise from interactions deeper in the rock than
those giving shield hits since both the muon and any associated charged
particles have to miss the shield.  Neutrons and photons produced
in these interactions would have to pass through more rock absorber.
The photon component would be attenuated faster than the neutron and the
resulting events would contain a reduced
fraction of shower events. 
\par The effect of absorption in the rock may be simulated by
calculating the track/shower ratio for different depth cuts in the main
detector. As in the rock,
the photon component is attenuated faster than the neutron
component.  The track/shower ratio, plotted as a function of depth cut (figure
\ref{ratio}(bottom)), rises to a plateau when the photon component
is completely absorbed.  We take $1.75\pm0.41$, the value at a depth of
80 cm, as our measurement of the ratio for pure
neutrons.  We estimate the systematic error on $R$ by
repeating the calculation with background ratios having values between those
measured for the full rock sample and the pure neutron sample, 
including allowance
for the errors on these numbers. This produces a
 variation of $R$ from 0.74 to 0.67. We take this variation
as an estimate of the systematic error.
Note that the possible rise in the background ratio due 
to absorption of the photon component
 results in a shift towards         
smaller R, i.e. further from the expected value of 1.0.
\par We have studied the effects of applying an extra cut on the depth
distribution to remove events closest to the exterior of the detector.  
Within the statistical limits on the data 
we see no significant change in $R$.  The uncut data
provides the maximum statistics and the best determination of the background
fraction and thus the smallest error on $R$.
\end{itemize}

\begin{table}
\caption[]{ Values of the components of the systematic error on R.\\}
\label{system}
\begin{tabular}{|l|c|}
\hline
      Error & $\delta R$ \\
\hline
Neutrino flux    & $\pm 0.038$ \\
Monte Carlo systematics & $\pm 0.03$ \\
Scanning systematics    & $\pm0.02$ \\
Background subtraction  & +0.02 -0.05 \\
    &  \\
\hline
Total systematic error  & +0.05 -0.07 \\
\hline
\end{tabular}
\end{table}
\vskip 0.5cm
\par The systematic errors are summarized in table \ref{system}.
\subsection{Absolute rates}
  The rate of track events is $0.79 \pm 0.18$
 of the expected rate and of showers $1.09 \pm 0.21$.  The errors do not
include the systematic error on the flux calculation, estimated to be $\pm20\%$
by reference \cite{Gaisser}.
If the BGS flux\cite{barr} is accurate we would support 
the hypothesis
that the anomaly results more from a loss of $\nu_\mu$ events than a gain
of $\nu_e$ events.  
\par  We have investigated other possible systematic effects on the absolute
rates.  These include uncertainties in the Fermi gas model, particularly
in the Pauli blocking of inelastic interactions producing a low energy
nucleon, uncertainties in the cross-sections,  
biases introduced by the detector trigger and biases in
the scanning process. We estimate that these produce a further 6.4\% 
systematic error on the ratio of measured to expected tracks and 7.5\% on
the showers.
\section{Conclusions}
\par   We have measured the flavour ratio of ratios ($R$) 
in atmospheric neutrino
interactions using a 1.52 kton-year exposure of Soudan~2.  
We find $R=0.72\pm0.19^{+0.05}_{-0.07}$.  This value is 
about $1.5\sigma$ from the expected value of 1.0 and is 
consistent with the anomalous ratios measured by the Kamiokande and IMB
experiments.  However we note that since our acceptance matrix is different 
from those of the water Cherenkov experiments we would not expect to measure
the same value of R, particularly
 if physics processes are occurring which are not simulated
in our Monte Carlo.  There is 
approximately a 7\% chance that our measurement would statistically give 
0.72 or less if the true answer is 1.0.  To this level we support the 
observation of an anomaly in the atmospheric neutrino flavour ratio 
in a detector using a completely different
detection technique and with different systematic biases.
Data taking in Soudan~2 is continuing and completion of our planned 5 
kton-year exposure in 1999 should definitively resolve the question of
the presence or otherwise of an anomaly.
\par We have investigated and corrected for backgrounds due to the interaction
of neutrons or photons produced by 
$\mu$ interactions in the rock surrounding
our detector. 
We have measured the track/shower ratio
for neutrons entering Soudan 2 and find a value of $1.75\pm0.41$.  Making
allowance for the fraction of the tracks (pions or protons) which would be
below Cherenkov threshold in water we cannot reduce this ratio significantly
below 1.0.  This is
in contradiction to the hypothesis \cite{Ryaz} that the
anomaly in the Kamiokande and IMB detectors could be due to a substantial
 excess of shower events in neutron background.

\begin{ack}  This work was undertaken with the support of 
the U.S. Department of Energy, the State and University of
Minnesota and the U.K. Particle Physics and
Astronomy Research Council.
 We wish to thank the following for their invaluable help with      
the Soudan 2 experiment: the staffs of the collaborating laboratories; the
Minnesota Department of Natural Resources for allowing us to use the facilities
of the Soudan Underground Mine State Park; the staff of the Park, particularly
Park Managers D. Logan and P. Wannarka, for their day to day support; and Messrs
B. Anderson, J. Beaty, G. Benson, D. Carlson, J. Eininger and J. Meier of the
Soudan Mine Crew for their work in the installation and running of the
experiment.
\end{ack}

\end{document}